\documentclass[final,3p,times]{elsarticle}

\usepackage{amssymb}
\usepackage{amsmath}
\usepackage{graphicx}
\usepackage{hyperref}

\biboptions{numbers,sort&compress}
\journal{Journal of Subatomic Particles and Cosmology}

\begin{document}

\begin{frontmatter}

\title{Experimental Review on Bulk Properties and Light/Strange Hadron Production in Heavy-Ion Collisions}

\author[ccnu]{Shusu Shi\corref{cor1}}
\author[thu]{Xianglei Zhu\corref{cor1}}

\cortext[cor1]{Corresponding authors: shiss@ccnu.edu.cn, zhux@tsinghua.edu.cn}

\affiliation[ccnu]{organization={Institute of Particle Physics, Central China Normal University},
             city={Wuhan},
             postcode={430079},
             country={China}}

\affiliation[thu]{organization={Department of Engineering Physics, Tsinghua University},
             city={Beijing},
             postcode={100084},
             country={China}}

\begin{abstract}
This proceedings article reviews recent experimental results on bulk properties and light/strange hadron production in heavy-ion collisions, presented at the Strangeness in Quark Matter 2026 conference. The discussion covers the QCD phase diagram, extraction of the temperature-dependent speed of sound, radial flow fluctuations probed by $v_0(p_T)$, anisotropic flow in small collision systems, strange hadron yields across system sizes, nuclear modification factors, and collectivity signatures at high baryon density. Results from the STAR, CMS, ALICE, ATLAS, PHENIX, NA61/SHINE, and HADES collaborations are synthesized to highlight recent progress in understanding the properties of the Quark-Gluon Plasma and the nature of the QCD phase transition.
\end{abstract}


\end{frontmatter}


\section{Introduction}

The study of bulk properties and hadron production in high-energy heavy-ion collisions provides critical insights into the behavior of strongly interacting matter under extreme conditions. The Quark-Gluon Plasma (QGP), a deconfined state of quarks and gluons, is believed to have existed in the early universe and can be recreated in laboratory experiments. At high beam energies corresponding to low net baryon density, experimental efforts focus on characterizing the properties of the QGP, including its equation of state, transport coefficients, and collective behavior. At lower energies where net baryon density becomes large, the primary interest shifts to mapping the QCD phase transition, searching for the possible critical point, and understanding the nature of matter in the high baryon density regime \cite{Luo:2020pef, Chen:2024aom, Chen:2026gka, He:2026fsq}.

Throughout this review, the term ``medium'' is used only as a generic shorthand. Where the physics permits, we distinguish explicitly between the partonic medium, i.e. the deconfined QGP with quark and gluon degrees of freedom, and the hadronic medium, i.e. the resonance-dominated hadron gas that follows hadronization. The relative weight of these two stages is governed by collision energy, system size, and baryon chemical potential: high-energy, low-\(\mu_B\) collisions form a long-lived partonic medium, whereas low-energy, high-\(\mu_B\) collisions are shorter-lived and more strongly influenced by hadronic interactions.

The QCD phase diagram is conventionally represented in the plane of temperature \(T\) and baryon chemical potential \(\mu_B\). At vanishing \(\mu_B\), lattice QCD calculations predict a smooth crossover from hadronic matter to QGP at a pseudo-critical temperature of approximately 156 MeV \cite{HotQCD:2018pds, Borsanyi:2020fev}. At high \(\mu_B\), various theoretical approaches suggest the existence of a first-order phase transition and a critical endpoint \cite{Fukushima:2010bq}. Mapping the phase diagram experimentally remains a primary objective in heavy-ion physics.
The experimental program at the Relativistic Heavy-Ion Collider (RHIC), the Large Hadron Collider (LHC), the Super Proton Synchrotron (SPS), and the Facility for Antiproton and Ion Research (FAIR) has produced a wealth of data spanning a wide range of collision energies and system sizes. In particular, the Beam Energy Scan (BES) program at RHIC has systematically explored the region of the QCD phase diagram at moderate to high baryon densities, collecting data at collision energies ranging from \(\sqrt{s_{NN}} = 7.7\) to 200 GeV, corresponding to \(\mu_B\) values from about 420 MeV down to 24 MeV. More recently, the BES-II program has extended the reach to even lower energies, including 3.0, 3.2, 3.5, 3.9, 4.5, and 5.2 GeV \cite{STAR:BES2whitepaper}.

In this context, the production of light and strange hadrons serves as a powerful probe of the collision dynamics. Since strange quarks are not present in the initial colliding nuclei, their production must occur during the evolution of the fireball and is therefore sensitive to the properties of the partonic medium and its subsequent hadronization. This article reviews recent experimental results on bulk properties and light/strange hadron production in heavy-ion collisions presented at the Strangeness in Quark Matter 2026 conference.

\section{Bulk Properties}

\subsection{Speed of Sound from Mean Transverse Momentum}

The speed of sound \(c_s^2(T)\) is a key thermodynamic quantity that characterizes the equation of state of strongly interacting QCD matter across the crossover.  It relates pressure variations to energy density changes and reflects the stiffness of the matter. In a conformal partonic system, \(c_s^2 = 1/3\), while in a hadron gas at low temperatures, \(c_s^2\) is substantially smaller. Lattice QCD calculations predict a rapid rise of \(c_s^2\) across the crossover transition.

The extracted \(c_s^2(T)\) characterizes the equation of state of the strongly interacting QCD medium across the crossover, with dominant sensitivity to the partonic stage and additional sensitivity to the crossover region. The extraction method relies on the relationship between \(\langle p_T \rangle\) and the expansion history of the fireball, which is sensitive to the equation of state. The extracted values show good agreement with Lattice QCD calculations, demonstrating the power of this new method to constrain the QGP properties directly from experimental data \cite{Broodo:SQM2026}.

\subsection{Radial Flow and Bulk Viscosity via \(v_0(p_T)\)}

Collective flow in heavy-ion collisions provides crucial information about the transport properties of the produced medium. While anisotropic flow coefficients \(v_n\) for \(n \geq 2\) are sensitive to the shear viscosity to entropy density ratio \(\eta/s\), the newly introduced observable \(v_0(p_T)\) is designed to probe radial flow fluctuations and bulk viscosity \(\zeta/s\). Unlike \(v_2\) or \(v_3\), which arise from initial state spatial anisotropies, \(v_0\) captures the event-by-event variation of the radial flow magnitude. Since radial flow develops through the partonic and hadronic stages, \(v_0(p_T)\) is sensitive to both the partonic transport properties and the subsequent hadronic expansion.

The ATLAS and ALICE collaborations have introduced \(v_0(p_T)\) as a novel probe of radial flow fluctuations \cite{ATLAS:2025ztg, ALICE:2025iud}. The STAR collaboration has presented a systematic study of the system-size dependence of \(v_0(p_T)\) at 200 GeV, demonstrating that this observable captures radial flow fluctuations with a coherent hydrodynamic response across different collision species \cite{Wang:SQM2026}.
STAR also has extended these measurements to multi-strange hadrons including \(\phi(1020)\), and \(\Xi\). The results show the characteristic mass ordering at low \(p_T\) (\(< 2\) GeV/\(c\)), where heavier particles have smaller \(v_0\) due to their larger inertia. At intermediate \(p_T\) (2-4 GeV/\(c\)), a clear baryon-meson splitting is observed between \(\Lambda\) and \(K_S^0\) at 200 GeV, consistent with the pattern expected from hydrodynamic flow. However, statistical precision at lower energies of 19.6 and 11.5 GeV remains limited, motivating further measurements at SPS, CBM and MPD~\cite{Kong:SQM2026}.

\subsection{Number of Constituent Quark Scaling}

The number of constituent quark (NCQ) scaling of elliptic flow \(v_2\) has long been considered a hallmark of partonic collectivity. When \(v_2\) is plotted as a function of transverse kinetic energy \((m_T - m_0)\) divided by the number of constituent quarks \(n_q\), particles of different masses and quark content tend to fall on a common curve. This scaling behavior is naturally explained if the flow develops at the partonic level before hadronization.

STAR observations show NCQ scaling in heavy-ion collisions at 200 GeV, consistent with the formation of a deconfined partonic medium where quark and gluon degrees of freedom dominate the collective expansion. The scaling has been verified for a wide range of particles including pions, kaons, protons, \(\Lambda\), \(\Xi\), \(\Omega\), and even \(D\) mesons at top RHIC and LHC energies \cite{STAR:2003wqp, STAR:2015gge, STAR:2017kkh, ALICE:2018yph}.
At lower beam energies, an interesting trend emerges from the BES program. NCQ scaling is found to be absent at the lowest BES energies but gradually restores as the collision energy increases from 3.2 to 4.5 GeV \cite{STAR:2013cow, STAR:2021yiu, STAR:2025owm, STAR:2025twg, STAR:2026axo}. This energy-dependent restoration provides evidence for the onset of partonic degrees of freedom in the collision system. The transition region around \(\sqrt{s_{NN}} \approx 4\) GeV may correspond to the energy where the system spends sufficient time in the deconfined partonic phase for partonic collectivity to develop, while at lower energies hadronic interactions dominate the expansion. More precise measurements of multi-strange hadron flow in future experiments at intermediate BES energies will help pinpoint the threshold for partonic behavior.

\section{Collectivity in Small Systems}

One of the most surprising and consequential developments in recent years has been the observation of anisotropic flow in small collision systems such as p+p, $p$+Au, $p$+Pb, $d$+Au, and $^3$He+Au. The systems were initially predicted to lack the scale required to sustain the hydrodynamic properties characteristic of large ion collisions. However, experimental results from the LHC and RHIC have demonstrated that collectivity persists even in these small systems, challenging our understanding of the necessary conditions for hydrodynamic behavior. Light-ion collisions, with well-constrained initial conditions from nuclear structure studies, can serve as a benchmark for hydrodynamic calculations in such small systems.

The CMS, ATLAS, ALICE, and STAR collaborations have all reported signatures of partonic collectivity in light-ion collisions \cite{Baty:SQM2026, Mohapatra:SQM2026, Machacek:SQM2026, Paul:SQM2026}. The observed elliptic flow \(v_2\) in O+O and Ne+Ne collisions exhibits many features previously thought to be unique to large systems, including mass ordering at low transverse momentum and baryon-meson splitting. Importantly, the baryon-meson splitting extends to higher transverse momentum at higher collision energies, consistent with the development of stronger radial flow in more energetic collisions. These observations indicate that a short-lived partonic medium can still generate sizable collective flow, although its lifetime and spatial extent are reduced compared with large-ion collisions.

The comparison between O+O and Ne+Ne collisions is particularly instructive. Both systems have similar mass numbers or sizes but different geometric properties. By studying the initial-state geometry dependence of flow observables, one can test predictions from hydrodynamic models and constrain the initial state fluctuations. The results demonstrate that partonic collectivity is robust across these small collision systems, with the measured flow coefficients scaling with the eccentricities of the respective initial geometries in line with hydrodynamic expectations.

The LHCb collaboration, using the SMOG2 fixed-target system, has also contributed novel results in this area. The anisotropic flow measurements in asymmetric Pb+Ne and Pb+Ar collisions provide higher sensitivity to the light nuclei deformation and the subsequent hydrodynamic response, since the initial-state eccentricities in central collisions are dominated by the light nucleus structure \cite{BoenteGarcia:2024kba, DaSilva:SQM2026}. These measurements thus offer complementary constraints on the mechanisms driving collective expansion in small systems.

\section{Light and strange hadron productions at low and intermediate baryon density}

\subsection{Nuclear Modification of Hadrons in Heavy-Ion Collisions}
Understanding how parton energy loss and nuclear modification evolve with collision energy and system size is a central goal of heavy-ion physics. When a high-energy parton traverses the deconfined partonic medium, it loses energy through gluon radiation and elastic scattering. This energy loss manifests as suppression of high transverse momentum hadrons, quantified by the nuclear modification factor \(R_{AA} = (dN/dp_T)_{AA} / [\langle N_{\text{coll}} \rangle (dN/dp_T)_{pp}]\) or the central-to-peripheral ratio \(R_{CP}\).

The nuclear modification factor for \(\phi(1020)\) mesons at forward rapidity in Au+Au collisions at 200 GeV, measured by the PHENIX collaboration, shows significant suppression at high transverse momentum, consistent with expectations from parton energy loss models \cite{PHENIX:SQM2026}. The \(\phi(1020)\) meson is particularly interesting because its small hadronic cross section with non-strange hadrons makes its suppression primarily sensitive to partonic energy loss rather than hadronic rescattering. This makes \(\phi(1020)\) a clean probe of the partonic-medium properties.

Beyond \(\phi(1020)\) mesons, the CMS collaboration has extended energy loss studies to smaller collision systems. Measurements of charged hadron \(R_{AA}\) in Ne+Ne and O+O collisions at \(\sqrt{s_{NN}} = 8.16\) TeV show significant suppression at high transverse momentum, with slightly stronger suppression observed in Ne+Ne than in O+O \cite{Pant:SQM2026}. This observation suggests an atomic mass number dependent suppression hierarchy, with larger systems showing stronger suppression as expected from the longer path length available for parton energy loss. However, the magnitude of the suppression in these light-ion systems is somewhat larger than simple scaling model predictions, indicating that the energy loss mechanism may have a non-trivial dependence on system size. These measurements provide important constraints on the fundamental parameters of parton energy loss, including the transport coefficient \(\hat{q}\) that characterizes the opacity of the partonic medium.

The comparison of \(R_{AA}\) and the central-to-peripheral ratio \(R_{CP}\) across different collision systems and energies enables systematic tests of energy loss models. 
At BES energies, the STAR collaboration has presented $R_{CP}$ measurements for strange hadrons and \(\phi(1020)\) mesons in Au+Au collisions from \(\sqrt{s_{NN}} = 7.7\) to 200 GeV. These measurements show that the high-$p_T$ suppression of strange hadrons and \(\phi(1020)\) mesons gradually weakens as the collision energy decreases and the baryon density increases, indicating a reduced partonic energy loss in the high baryon density region \cite{Yuan:SQM2026}.
These results help constrain the energy dependence of parton energy loss and its relationship to the properties of the produced partonic medium.

\subsection{The \(\Omega/\phi(1020)\) Ratio in BESII: Evidence for the Onset of Deconfinement}

The ratio of the yield of \(\Omega\) baryons to that of \(\phi(1020)\) mesons provides a sensitive probe of the strange quark production mechanism and the nature of the medium created in heavy-ion collisions, in particular whether it is partonic or hadronic. The \(\Omega\) baryon (sss) contains three strange quarks, while the \(\phi(1020)\) meson (s\(\bar{s}\)) contains one strange quark-antiquark pair. Consequently, the production of \(\Omega\) relative to \(\phi(1020)\) is expected to differ significantly between hadronic and partonic scenarios. In the partonic scenario, the \(\Omega\) is produced through coalescence of strange quarks, which enhances its yield relative to the \(\phi(1020)\) at intermediate \(p_T\), whereas in a purely hadronic scenario the \(\Omega\) must be produced through multi-step hadronic reactions that are strongly suppressed. This makes the \(\Omega/\phi(1020)\) ratio a potential diagnostic for the onset of deconfinement.

The STAR collaboration has presented measurements of the \(\Omega/\phi(1020)\) ratio as a function of $p_T$ in the BES-II program \cite{Yuan:SQM2026}. These measurements are performed in Au+Au collisions spanning a range of \(\sqrt{s_{NN}}\) from 7.7 to 19.6 GeV. 
The data show a strong enhancement of the \(\Omega/\phi\) ratio at intermediate \(p_T\) towards central collisions at all energies.
To interpret these results, comparisons have been made with the AMPT (A Multi-Phase Transport) model. Two versions of the AMPT model are considered: the default version, which uses a hadronic transport approach without partonic coalescence, and the string melting version, which includes partonic degrees of freedom and coalescence for hadronization. The string melting AMPT model, which incorporates a deconfined partonic medium, is found to describe the measured \(\Omega/\phi(1020)\) ratio well for central Au+Au collisions at \(\sqrt{s_{NN}} \geq 7.7\) GeV. In contrast, the default AMPT model, which does not include partonic dynamics, significantly underestimates the \(\Omega/\phi(1020)\) ratio at these energies.

This comparison leads to an important conclusion: the onset of deconfinement in central Au+Au collisions occurs at \(\sqrt{s_{NN}} \geq 7.7\) GeV. The agreement between the string melting AMPT model and the experimental data indicates that partonic degrees of freedom become relevant in this energy regime, enabling enhanced production of multi-strange baryons through parton coalescence. The relative flatness of the \(\Omega/\phi(1020)\) ratio across the BES energy range is consistent with the interpretation that once deconfinement is established above 7.7 GeV, the strangeness production mechanism remains dominated by partonic processes, with no dramatic changes in the \(\Omega/\phi(1020)\) ratio as the collision energy increases further.
Future high-statistics measurements of the \(\Omega/\phi(1020)\) ratio at intermediate BES energies, particularly near the threshold region, will help refine our understanding on the onset of deconfinement. The combination of \(\Omega/\phi(1020)\) with other strange hadron ratios (such as $K^+/\pi^+$ \cite{Lewicki:SQM2026}) and flow observables will provide a comprehensive picture of the strangeness production dynamics across the QCD phase diagram.

\subsection{Excess of Charged over Neutral Kaons}

NA61/SHINE has reported an excess of charged over neutral kaons in Ar+Sc collisions at 8.8 GeV \cite{Lewicki:SQM2026}. This observation confirms earlier measurements and points to interesting isospin effects in strangeness production at high baryon density. The excess may arise from differences in the production and re-scattering of charged and neutral kaons in the dense hadronic medium. Alternatively, it could be a signature of the formation of a deconfined partonic phase, where the production of strange quarks favors the creation of charged strange mesons. Further measurements across different collision systems and energies are needed to distinguish between these possibilities.

\section{Strangeness Production at High Baryon Density}

The region of the QCD phase diagram at high baryon density remains relatively unexplored, and recent results from the BES program have begun to shed light on the properties of matter in this regime. Strangeness production is of particular interest because strange quarks are not present in the initial colliding nuclei; their production must occur during the collision and is sensitive to the properties of the dense, baryon-rich medium, where hadronic interactions are expected to play a dominant role.

\subsection{Yields and Centrality Scaling}

Strange hadron yields scale smoothly with the number of participants \(\langle N_{\text{part}} \rangle\) across different collision systems, from small systems like p+p to large systems like Au+Au \cite{Ponce:SQM2026}. This smooth scaling indicates a universal production mechanism, with the yield per participant increasing in more central collisions due to the larger volume and longer lifetime of the fireball.

At high baryon density, the canonical ensemble description provides a better fit to strange hadron yields than the grand-canonical ensemble \cite{STAR:2024znc}. In the canonical ensemble, strangeness is conserved exactly, leading to suppression of strange particle yields in small systems where the number of produced strange quarks is small. The preference for the canonical ensemble at high baryon density suggests that strangeness conservation is enforced globally rather than locally, consistent with the smaller volume and shorter lifetime of fireballs created in low-energy collisions, where the hadronic medium dominates the late-stage evolution.

The centrality scaling parameter \(\alpha\), defined by \(Y = k \langle N_{\text{part}} \rangle^\alpha\), provides additional insight into the production mechanism. For \(\Xi^-\), the first near threshold measurement in Au+Au collisions reveals an \(\alpha\) value significantly larger than that for \(\Lambda\), \(K_S^0\), and \(\phi(1020)\) \cite{Labonte:SQM2026}. This enhanced centrality dependence suggests that multi-strange baryons are particularly sensitive to the volume and density of the hadronic medium, with hadronic interactions dominating their production in the high baryon density region.

\subsection{Directed Flow of \(\phi(1020)\) Meson}

The directed flow \(v_1\) of the \(\phi(1020)\) meson at beam energies of 3 to 4.5 GeV provides particularly intriguing insights into the properties of high baryon density matter. Directed flow, which describes the sideward deflection of particles in the reaction plane, is sensitive to the early-stage pressure gradients and the equation of state at high density.

While the UrQMD model with mean-field reproduces proton flow data, it fails to describe the \(\phi(1020)\) flow unless high-mass baryon resonances are included \cite{Zheng:SQM2026}. This observation implies that high-mass baryon resonances (\(N^*\)) may play a crucial role in driving \(\phi(1020)\) production and its baryon-like flow pattern. The \(\phi(1020)\) meson has an unusually small cross section with nucleons, allowing it to escape the fireball with relatively little re-scattering. Consequently, its flow pattern reflects the early-stage dynamics more faithfully than that of other hadrons. The fact that \(\phi(1020)\) exhibits strong flow while other mesons do not may be explained by the small \(\phi(1020)\)-\(N\) cross section, which allows resonance decays to dominate \(\phi(1020)\) production and imprint baryon-like flow on the \(\phi(1020)\) meson.

These results highlight the importance of understanding resonance production and decay in the high baryon density region, where the medium is hadron dominated. Improved models that include a complete set of high-mass resonances will be necessary for quantitative interpretation of the data.

\subsection{Hyperonic Thermometer}

The HADES collaboration has reported a \(\Lambda/\Sigma\) ratio of \(3.1 \pm 0.8\) in high baryon density collisions \cite{Orlinski:SQM2026}. This ratio provides a new thermometer for the dense hadronic medium because the production of these hyperons is sensitive to the temperature and baryon chemical potential of the fireball. Using a thermal model, the measured ratio can be converted into a temperature, yielding an estimate of the kinetic freeze-out temperature in the high baryon density regime. This "hyperonic thermometer" complements other freeze-out temperature determinations from slope parameters and particle ratios, providing a consistent picture of the late-stage evolution of the fireball when hadronic degrees of freedom dominate.

\section{Summary and Outlook}

Recent experimental results from RHIC, LHC, SPS, and GSI have significantly advanced our understanding of bulk properties and strangeness production in heavy-ion collisions. A central theme emerging from these measurements is the need to distinguish the partonic and hadronic stages of the produced medium and to understand how their relative importance depends on collision energy, system size, and baryon chemical potential.

At high collision energies and low \(\mu_B\), the fireball spends a substantial time in a deconfined partonic medium. The properties of this partonic medium are reflected in the first extraction of the temperature-dependent speed of sound, which agrees with lattice QCD, in the anisotropic and radial flow that constrain \(\eta/s\) and \(\zeta/s\), in the NCQ scaling of \(v_2\), and in the high-\(p_T\) suppression of hadrons due to parton energy loss. Collectivity in this regime develops at the partonic level, and the observation of hydrodynamic-like signals in small systems such as O+O, Ne+Ne, \(p\)+Pb, \(d\)+Au, and \(^3\)He+Au shows that a short-lived partonic medium can still generate sizable collective flow, although with reduced lifetime and stronger initial-state fluctuations than in large-ion collisions.

At lower collision energies and higher \(\mu_B\), the partonic stage becomes shorter and the hadronic medium plays an increasingly important role. This change is visible in the energy dependence of NCQ scaling, which is absent at the lowest BES energies and gradually restored from 3.2 to 4.5 GeV; in the enhanced \(\Omega/\phi(1020)\) ratio at \(\sqrt{s_{NN}} \geq 7.7\) GeV, which is described by the string-melting AMPT model but not by the default hadronic-transport version; in the canonical suppression of strange hadron yields; in the enhanced centrality dependence of multi-strange baryons; and in the \(\phi(1020)\) directed flow, which requires high-mass baryon resonances and points to resonance-dominated hadronic dynamics. The \(\Lambda/\Sigma\) ratio further provides a hyperonic thermometer for the dense, baryon-rich hadronic medium at freeze-out.

System-size dependence provides a complementary handle on this global picture. In large systems, the longer lifetime and larger volume allow partonic collectivity and parton energy loss to develop fully, leading to clear NCQ scaling, mass ordering, and high-\(p_T\) suppression. In small systems, partonic-like collectivity persists at high energy, but the smaller volume and shorter lifetime enhance initial-state fluctuations and reduce the time available for partonic and hadronic rescattering. At high baryon density, the smaller fireball volume also makes exact strangeness conservation more relevant, favoring canonical-ensemble descriptions of strange hadron production. Thus, the transition from partonic-dominated to hadron-dominated dynamics is governed not only by collision energy but also by system size.

The first extraction of the temperature-dependent speed of sound from experimental data demonstrates the power of sophisticated analysis techniques to constrain the QCD equation of state. The introduction of \(v_0(p_T)\) as a probe of radial flow fluctuations and bulk viscosity opens a new window into the transport properties of the QGP and the crossover region. The persistence of partonic collectivity in small collision systems challenges our understanding of the necessary conditions for hydrodynamic behavior and motivates further theoretical developments.

Energy loss studies have revealed a clear system-size dependence, with the suppression pattern in Ne+Ne and O+O collisions providing constraints on the fundamental parameters of parton energy loss. The \(\Omega/\phi(1020)\) ratio measured in BESII provides evidence for the onset of deconfinement in Au+Au collisions at \(\sqrt{s_{NN}} \geq 7.7\) GeV. At high baryon density, hadronic interactions appear to dominate particle production, and the canonical ensemble provides a better description of strange yields than the grand-canonical ensemble. The directed flow of \(\phi(1020)\) mesons points to the important role of high-mass resonances, and the \(\Lambda/\Sigma\) ratio offers a new thermometer for dense hadronic matter.

The BES-II program provides higher statistics data at multiple collision energies, enabling more precise measurements of strange hadron flow and production. Future measurements below 7.7 GeV will be particularly important for refining our understanding of the deconfinement onset. In addition to RHIC BES-II, other experimental programs will play crucial roles in exploring the high baryon density region. The CSR External-target Experiment (CEE) at HIRFL-CSR, the Multi-Purpose Detector (MPD) at the Nuclotron-based Ion Collider fAcility (NICA), and the Compressed Baryonic Matter (CBM) experiment at FAIR will provide complementary coverage of the QCD phase diagram at high baryon chemical potentials~\cite{Lu:2016htm, Liu:2023xhc, MPD:2025jzd, CBM:2025voh}. These facilities, together with the ongoing programs at SPS and RHIC, will systematically explore the phase structure of strongly interacting matter and search for the QCD critical point. Together, these efforts will continue to refine our understanding of the QCD phase diagram and the properties of strongly interacting matter under extreme conditions.

\section*{Acknowledgments}

This work is supported in part by the National Key Research and Development Program of China
under Grant nos. 2022YFA1604900 and 2024YFA1610700, and the Fundamental Research Funds for the Central Universities(XJ2026000302).

\bibliographystyle{elsarticle-num}
\bibliography{sqm2026_template}

\end{document}